\documentclass[iop,twocolappendix]{emulateapj}
\usepackage{amsmath}
\usepackage[cal=boondox,calscaled=0.95,bb=px]{mathalfa} 

\newcommand{\lux}{\texttt{lux}}
\newcommand{\gray}{\texttt{GRay}}
\newcommand{\grayii}{\texttt{GRay2}}
\newcommand{\srh}{\vphantom{$\sqrt{2}$}}

\begin{document}

\title{\grayii: A General Purpose Geodesic Integrator for Kerr Spacetimes}

\author{%
  Chi-kwan Chan\altaffilmark{1,2},
  Lia Medeiros\altaffilmark{1,3,2},
  Feryal \"Ozel\altaffilmark{1,2,*}, and
  Dimitrios Psaltis\altaffilmark{1,4,2}
}

\altaffiltext{1}{Steward Observatory and Department of Astronomy,
  University of Arizona,
  933 N. Cherry Ave., Tucson, AZ 85721}

\altaffiltext{2}{Black Hole Initiative, Harvard University, 20 Garden St.,
  Cambridge, MA 02138}

\altaffiltext{3}{Department of Physics,
  Broida Hall, University of California Santa Barbara,
  Santa Barbara, CA 93106}

\altaffiltext{4}{Radcliffe Institute for Advanced Study,
  Harvard University, Cambridge, MA 02138, USA}

\altaffiltext{*}{Guggenheim Fellow}

\begin{abstract}
  Fast and accurate integration of geodesics in Kerr spacetimes is an
  important tool in modeling the orbits of stars and the transport of
  radiation in the vicinities of black holes.
  Most existing integration algorithms employ Boyer-Lindquist
  coordinates, which have coordinate singularities at the event
  horizon and along the poles.
  Handling the singularities requires special numerical treatment in
  these regions, often slows down the calculations, and may lead to
  inaccurate geodesics.
  We present here a new general-purpose geodesic integrator, \grayii,
  that overcomes these issues by employing the Cartesian form of
  Kerr-Schild coordinates.
  By performing particular mathematical manipulations of the geodesic
  equations and several optimizations, we develop an implementation of
  the Cartesian Kerr-Schild coordinates that outperforms calculations
  that use the seemingly simpler equations in Boyer-Lindquist coordinates.
  We also employ the \texttt{OpenCL} framework, which allows
  \grayii\ to run on multi-core CPUs as well as on a wide range of GPU
  hardware accelerators, making the algorithm more versatile.
  We report numerous convergence tests and benchmark results for
  \grayii\ for both time-like (particle) and null (photon) geodesics.
\end{abstract}


\section{Introduction}

Integrating geodesics of particles and photons in the spacetimes of
Kerr black holes is an important aspect of theoretical modeling of
various astrophysical phenomena, from the orbits of stars and compact
objects around supermassive black holes~\citep[see,
  e.g.,][]{Alexander2017} to the transport of radiation through their
accretion flows~\citep[see, e.g.,][]{Yuan2014}.
Fast geodesic integrators are also critical in fitting data of, e.g.,
stars in orbit around the black hole in the center of the Milky
Way~\citep{Boehle2016,Gillessen2017}, of rotationally broadened
fluorescence lines from accreting black holes in the
X-rays~\citep{Miller2007}, or of interferometric data taken with the
Event Horizon Telescope that aims to take the first image of a
supermassive black holes with horizon scale resolution~\citep[see,
  e.g.,][]{2008Natur.455...78D, Doeleman2012}.

Calculations of test-particle orbits (time-like geodesics) around
black holes has been traditionally done in a post-Newtonian
approximation, focusing on N-body effects~\citep[see, e.g.,][for
  recent work]{Brem2014, Hammers2014}, or by solving simultaneously
for the dynamical spacetime of the cluster of
particles~\citep[see][and references therein]{Shapiro1992}.
More recently, fast algorithms have been developed that follow the
orbits of test particles in stationary black-hole spacetimes, with no
approximations~\citep{Yang2014,Zhang2015}.

Integrations of null geodesics (ray tracing) in Kerr spacetimes can be
traced back to \citet{Bardeen1973}, \citet{Cunningham1975}, and
\citet{1979A&A....75..228L}, where the images of accretion disks and
the outlines of the shadows of Schwarzschild and extreme Kerr black
holes were first obtained.
More recently, methods of combining polarized radiative transfer with
ray tracing \citep[see, e.g.,][]{2003MNRAS.342.1280B,
  2004MNRAS.349..994B, Gammie2012, Younsi2012, 2013ApJ...777...11S} as
well as a variety of open-source algorithms for fast radiative
transfer calculations have been developed~\citep[see,
  e.g.,][]{2009ApJ...696.1616D,2009ApJS..184..387D,
  2013ApJ...777...13C, 2013ApJS..207....6Y, 2016ApJ...820..105P,
  Dexter2016}.

In an earlier article, we described \gray~\citep{2013ApJ...777...13C},
the first publicly available numerical algorithm that made explicit
use of general-purpose computing on graphics processing units (GPU)
for ray tracing in relativistic spacetimes.
\gray\ uses the high computational horsepower of GPUs to speed up this
computationally intensive problem.
It achieved 1--2 orders of magnitude speed up compared traditional
CPU-base algorithms and allowed us to generate large, high-cadence
simulations of the observable properties of accreting black holes
\citep{2015ApJ...799....1C, 2015ApJ...812..103C, 2015ApJ...814..115P,
  2016ApJ...832..156K, 2016ApJ...826...77B, 2016arXiv161003505M,
  2016arXiv160106799M}.

Even though \gray\ is very fast and efficient, it uses a standard
physical setup of the ray-tracing problem as well as numerical methods
that have a number of limitations.
For example, like most of the other algorithms, \gray\ employs the
Boyer-Lindquist (BL) coordinates to take advantage of the symmetry of
the Kerr spacetime, which greatly simplifies the derivation and
evaluation of the Christoffel symbols.
However, the various coordinate singularities in the BL coordinates
cause numerical difficulties.
Moreover, as in many other algorithms, \gray\ uses the so called
fast-light approximation \citep[see, however,][]{2009ApJS..184..387D}.
This means that, when solving the radiative transfer equation along
each ray, the fluid is assumed to be time independent, or
equivalently, the speed of each photon is taken to be effectively
infinite.
This approximation greatly simplifies the algorithms because only a
single snapshot of the underlying matter through which radiation
propagates is needed in the radiative transfer calculation at each
time step.
However, this assumption affects the time variability properties of
the simulations at the fastest timescales near the black-hole horizons
\citep{2009ApJS..184..387D}, which will be important in interpreting
the upcoming observations with the Event Horizon
Telescope~\citep{2016ApJ...832..156K, 2016arXiv161003505M,
  2016arXiv160106799M}.

In order to overcome these difficulties, we describe here \grayii, a
new open source, hardware accelerated, geodesic integration algorithm.
We improve the geodesic integration in \grayii\ by switching to the
Cartesian form of Kerr-Schild (KS) coordinates, overcoming all
coordinate singularities and increasing the overall accuracy of the
calculations.
We also switch to \texttt{OpenCL}, which allows \grayii\ to run on
multi-core CPUs as well as on a wide range of hardware accelerators,
making the algorithm more versatile.
Finally, \grayii\ can handle both time-like (test particle) and null
(photon) geodesics, making it applicable to calculations of both
stellar orbits and radiative transfer.

In the next section, we discuss the limitations of using the BL
coordinates and derive an optimized form of geodesic equations in the
Cartesian KS coordinates.
In section~\ref{sec:ct}, we provide the details of using coordinate
time instead of the affine parameter to integrate the geodesic
equations.
In section~\ref{sec:implementation}, we summarize the implementation
details of \grayii.
In section~\ref{sec:tests}, we perform a convergence study using
unstable spherical photon orbits and stable particle orbits.
In section~\ref{sec:benchmarks}, we report benchmark results of
\grayii\ running on a wide range of CPUs and GPUs and demonstrate that
GPUs can be up to two orders of magnitude faster than a single CPU
core and that integrating in Cartesian KS coordinates can outperform
integrating in BL coordinates.
Finally, we summarize our findings in section~\ref{sec:summary}.

\section{The \grayii\ Algorithm}

\subsection{Implementation of the Cartesian Kerr-Schild Coordinates}
\label{sec:KS}

Letting $M$ and $a$ be the mass and spin parameter of a Kerr black
hole, the Boyer-Lindquist line element reads\footnote{We use script
  symbols $(\mathcal{t}, \mathcal{r}, \vartheta, \varphi)$ to denote
  the BL coordinates.
  Standard italic symbols $(t, r, \theta, \phi)$ and $(t, x, y, z)$
  are reserved for the spherical polar and Cartesian forms of the KS
  coordinates.}
\begin{align}
  ds^2 &=
  - \left(1 - \frac{2M\mathcal{r}}{\varrho^2}\right) d\mathcal{t}^2
  - \frac{4M\mathcal{r}a\sin^2\!\vartheta}{\varrho^2}d\varphi d\mathcal{t}
  + \frac{\varrho^2}{\varDelta}d\mathcal{r}^2 \nonumber\\
  &\!\!\!+ \varrho^2 d\vartheta^2
  + \left(\mathcal{r}^2 + a^2 + \frac{2M\mathcal{r}a^2\sin^2\!\vartheta}%
  {\varrho^2}\right)\sin^2\!\vartheta d\varphi^2,
\end{align}
where $\varrho^2 \equiv \mathcal{r}^2 + a^2\cos^2\vartheta$ and
$\varDelta \equiv \mathcal{r}^2 - 2M\mathcal{r} + a^2$.
The spherical polar nature of the BL coordinates introduces
unnecessary coordinate singularities along the poles, in addition to
the coordinate singularities at the event horizons.
When integrating geodesics numerically, these coordinate singularities
can cause significant difficulties.
Although there exist algorithms to overcome these difficulties
\citep[see, e.g., the method introduced by][]{2013ApJ...777...13C},
the poles can still cause numerical problems such as slowing down the
calculations, leading to inaccurate geodesics, and even crashing the
algorithms for extreme cases such as computing a face-on image of a
black hole accretion disk at inclination $i = 0^\circ$.

In \grayii, we resolve the coordinate singularities by employing the
Cartesian form of the KS coordinates
\begin{align}
  g_{\alpha\beta} &= \eta_{\alpha\beta} + f l_\alpha l_\beta, \label{def:CKS}
\end{align}
where $\eta_{\alpha\beta} \equiv \mathrm{diag}(-1,1,1,1)$ is the
Minkowski metric,
\begin{align}
  f        &= \frac{2r^3}{r^4 + a^2 z^2}, \\
  l_\alpha &= \left(1,
                    \frac{rx + ay}{r^2 + a^2},
                    \frac{ry - ax}{r^2 + a^2},
                    \frac{z}{r}\right),
\end{align}
and $r$ is defined implicitly by
\begin{align}
  x^2 + y^2 + z^2 = r^2 + a^2\left(1 - z^2 / r^2\right).
  \label{eq:CKSr}
\end{align}
Its Cartesian nature not only completely avoids the coordinate
singularities along the poles but requires no special treatment in the
integrator.
This is an advantage for implementing numerical integrators on modern
hardware accelerators, such as GPUs, because these massive parallel
stream processors use the Single-Instruction Multiple-Data (SIMD)
paradigm, which are inefficient in handling branch instructions (i.e.,
conditional statements).

Besides the poles, both the spherical and Cartesian KS coordinates are
also horizon-penetrating---there is no coordinate singularity at the
event horizons.
We can, in principle, integrate geodesics through the event horizon
into the interior of the black hole.
In fact, this property makes the spherical polar form of KS
coordinates the default choice for many GRMHD codes \citep[see,
  e.g.,][]{2003ApJ...589..444G, 2009ApJ...704..937N,
  2013MNRAS.429.3533S, 2015ApJ...807...31R, 2016ApJS..225...22W}, as
no special boundary treatment is required at the horizons.

Given that all the elements are non-zero in the Cartesian KS metric,
this may seem at first to be a computationally very expensive
coordinate system to work with.
A rough operation-count goes as following.
For the BL coordinates, there are 5 independent, non-zero elements in
the metric, 10 independent elements in the metric derivative tensor
$g_{\mu\nu,\alpha}$, and 20 independent Christoffel symbols.
In contrast, in the Cartesian KS coordinates, we need to compute all
10 independent metric elements.
The metric is time-independent but does not use any spatial symmetry,
resulting in 30 independent elements in the metric derivative tensor.
Since the metric derivative tensor tends to be the most complicated
part of the calculation, this suggests that the computation of the 40
Christoffel symbols in the KS coordinates requires roughly 3 times
more operations than in the BL coordinates.
Therefore, we expect solving the geodesic equations in the Cartesian
KS coordinates to be at least 3 times more expensive than in the BL
coordinates.

For each geodesic, we need to solve all four second-order ordinary
differential equations, if we integrate with respect to the affine
parameter $\lambda$.
Comparing to the methods that use the Killing vectors \citep[see,
  e.g.,][]{2012ApJ...745....1P, 2013ApJ...777...13C}, this is a
\begin{align}
  \frac{2\times(8~\mbox{variables})}
       {2\times(6~\mbox{variables}) + (1~\mbox{constant})} - 1 = 23\%
  \label{eq:bandwidth}
\end{align}
increase in the bandwidth requirement.
Nevertheless, as we will show in our benchmarks, geodesic integration
is in general \emph{compute-bounded}, meaning that the performance is
limited by the speed of the computation and not by the speed of
transferring data.
Alternatively, we can also integrate the geodesic equations with
respect to the coordinate time $t$ (see next section).
This way, the number of dynamic variables reduces to six, which is the
same as for \texttt{GRay} \citep{2013ApJ...777...13C}.

One of the important improvements in \grayii\ is that, by a series of
mathematical manipulations and regrouping, we significantly reduce the
operation-count of the geodesic equations in the Cartesian KS
coordinates.
Let $\lambda$ be the affine parameter and $\dot{x}^\mu \equiv
dx^\mu/d\lambda$.
Our manipulations start by realizing that, although the Christoffel
symbols $\varGamma^\mu_{\alpha\beta}$ provide an elegant form of
writing the geodesic equations
\begin{align}
  \ddot{x}^\mu
  &= -\varGamma^\mu_{\alpha\beta} \dot{x}^\alpha \dot{x}^\beta,
\end{align}
it is actually more efficient to go back a step and write the
equations in terms of the metric derivative tensor as
\begin{align}
  \ddot{x}^\mu
  &= -\frac{1}{2} g^{\mu\nu}(g_{\nu\alpha,\beta}
                           + g_{\nu\beta,\alpha}
                           - g_{\alpha\beta,\nu}) \dot{x}^\alpha \dot{x}^\beta
                           \label{eq:geo} \\
  &= -g^{\mu\nu} g_{\nu\alpha,\beta} \dot{x}^\alpha \dot{x}^\beta
     +\frac{1}{2} g^{\mu\nu} g_{\alpha\beta,\nu} \dot{x}^\alpha \dot{x}^\beta.
     \label{eq:nosym}
\end{align}
We can combine the first two terms in equation~(\ref{eq:geo}) because
the product of an anti-symmetric tensor and a symmetric tensor
vanishes---there is no need to explicitly symmetrize $\alpha$ and
$\beta$ for $g_{\nu\alpha,\beta}$ from a computational point of view.
Furthermore, equation~(\ref{eq:nosym}) can be written as following by
replacing the indices
$\nu\rightarrow\beta\rightarrow\alpha\rightarrow\gamma$ for the first
term and $\nu\rightarrow\alpha\rightarrow\beta\rightarrow\gamma$ for
the second term
\begin{align}
  \ddot{x}^\mu
  &= -\left(g^{\mu\beta}\dot{x}^\alpha -
            \frac{1}{2}g^{\mu\alpha}\dot{x}^\beta\right)
     g_{\beta\gamma,\alpha}\dot{x}^\gamma.
  \label{eq:generic}
\end{align}
For each geodesic, $\dot{x}^\mu$ depend only on $\lambda$.
Although we still need to evaluate all 30 independent non-zero
elements of $g_{\beta\gamma,\alpha}$, we can store their results into
the 12 non-zero elements the term outside the parenthesis in the above
equation and reuse them in the summation of each $\mu$.
Therefore, even in this general form without specifying a metric, the
geodesic equations are in fact simpler than how they look, at least in
terms of operation-count.

Next, by substituting the definition of the Cartesian KS
metric~(\ref{def:CKS}) into equation~(\ref{eq:generic}), we obtain
\begin{align}
  \ddot{x}^\mu
  &= - \left(\eta^{\mu\beta} \dot{x}^\alpha -
             \frac{1}{2}\eta^{\mu\alpha} \dot{x}^\beta\right)
     \dot{x}_{\beta,\alpha} + F l^\mu
\end{align}
with
\begin{align}
  F &\equiv f \left(l^\beta \dot{x}^\alpha -
               \frac{1}{2}l^\alpha \dot{x}^\beta\right)
       \dot{x}_{\beta,\alpha}.
\end{align}
In the above equations, the Minkowski metric $\eta^{\mu\nu}$
effectively picks out different components of the derivative tensor
and applies different signs.
Hence, we can split the equations and optimize them further as
\begin{align}
  \ddot{x}^0 &=   \dot{x}^\alpha \dot{x}_{0,\alpha} - F,
    \label{eq:opt0}\\
  \ddot{x}^i &= - \dot{x}^\alpha
    \left(\dot{x}_{i,\alpha} - \frac{1}{2}\dot{x}_{\alpha,i}\right) + F l^i.
    \label{eq:opti}
\end{align}
Note that the positions of the indices 0 and $i$ do not match on the
two sides of the above equations.
This is not an error; equations~(\ref{eq:opt0}) and (\ref{eq:opti})
are no longer tensor equations.

In the above new form, the right hand sides (RHS) of the geodesic
equations in the Cartesian KS coordinates have only $\sim 65\%$ more
floating-point operations than in the BL coordinates.
This is less than half of the operations compared to our rough
estimate.
Furthermore, the evaluation of the RHS uses many matrix-vector
products, which are optimized in modern hardware.
Indeed, as we show below, our benchmarks (Table~\ref{tab:elapse} and
\ref{tab:speedup}) show that using the Cartesian KS on discrete GPUs
can outperform its BL counterpart.

\subsection{Coordinate Time Integration}
\label{sec:ct}

In order to efficiently overcome the fast-light approximation, we need
to control the integration of a geodesic to a targeted time according
to the GRMHD simulations, which are usually performed in the spherical
KS coordinates \citep[see, e.g.,][]{2003ApJ...589..444G,
  2013MNRAS.429.3533S}.
This minimizes both the data reading and memory overhead by requiring
only sequential reading with at most two snapshots in memory.
In addition, we can take advantage of the fact that GPUs have special
purpose hardware for accelerating interpolation, which makes accessing
GRMHD simulations essentially free of overhead.

In \grayii, we develop two classes of methods to integrate the
geodesic equations to a target KS coordinate time: (\emph{i})~by
directly integrating the geodesics with respect to the KS coordinate
time and (\emph{ii})~by applying different root finders to the
numerical solutions to match the targeted time\footnote{These two
  methods are not mutually exclusive.
  In principle, we can combine them to integrate with respect to one
  coordinate and match a target value in another coordinate or variable.
  This may be useful for performing, e.g., Monte Carlo scattering
  simulations.}.
In this section, we will limit our discussion to method~(\emph{i}) and
derive the geodesic equations in terms of the KS coordinate time.
Again, these equations are used in \grayii\ mainly to overcome the
fast-light approximation.
Although they also reduce the required bandwidth of the geodesic
integrator, the performance impact is very minor.

Letting $v^\mu \equiv dx^\mu/dt$ and using the chain rule, it is
straightforward to derive the geodesic equations in the following form
\begin{align}
  \frac{dv^i}{dt} &= -\varGamma^i_{\alpha\beta} v^\alpha v^\beta
                     +\varGamma^0_{\alpha\beta} v^\alpha v^\beta v^i.
\end{align}
Note that we only consider the spatial components of the geodesic
equations.
The time component $dv^0/dt = 0$ is trivial and consistent with the
spatial part of the equations \citep[see,
  e.g.,][]{1993tegp.book.....W}.
Substituting the definition of the Christoffel symbols, we get
\begin{align}
  \frac{dv^i}{dt}
  &= -\frac{1}{2} g'^{i\nu}(g_{\nu\alpha,\beta}
                          + g_{\nu\beta,\alpha}
                          - g_{\alpha\beta,\nu}) v^\alpha v^\beta,
                          \label{eq:geo_t}
\end{align}
where $g'^{i\nu} \equiv g^{i\nu} - v^i g^{0\nu}$.
Equations~(\ref{eq:geo_t}) and (\ref{eq:geo}) have the same form.
Hence, the optimizations we carried out in the last section are still
applicable; equation~(\ref{eq:geo_t}) can be optimized to the generic
form
\begin{align}
  \frac{dv^i}{dt}
  &= -\left(g'^{i\beta}v^\alpha - \frac{1}{2}g'^{i\alpha}v^\beta\right)
     g_{\beta\gamma,\alpha}v^\gamma.
\end{align}
Finally, substituting the definition of the Cartesian KS metric, we
obtain
\begin{align}
  \frac{dv^i}{dt}
  &= - v^\alpha\left(v_{i,\alpha} - \frac{1}{2}v_{\alpha,i}\right) + F l^i
  \nonumber\\[-6pt]
  &\hspace{102pt} - (v^\alpha v_{0,\alpha} - F) v^i.
\end{align}

The first two terms in the RHS of the above equation (i.e., first
line) match the RHS of equation~(\ref{eq:opti}), while the last term
(i.e., second line) matches the RHS of equation~(\ref{eq:opt0}).
This is expected and in fact shows that integrating with respect to
the affine parameter and coordinate time have the same computational
complexity.
Therefore, because numerically integrating geodesics is compute
bounded, the choice between integrating with respect to the affine
parameter or with respect to the coordinate time will depend on the
application of \grayii\ and not on the detailed performance of each
method.
If one cares only about calculating the shapes of geodesics, then
using coordinate time will reduce the number of variables and save
some bandwidth as shown by equation~(\ref{eq:bandwidth}).
If, on the other hand, the radiative transfer equation needs to be
integrated along a geodesic, then using the affine parameter will give
the gravitational redshift with no additional
computations\footnote{Note that we can use the constant of motion
  $v^\alpha v_\alpha$ to solve for $v^0$ from $v^i$.
  In such a case, however, $v^\alpha v_\alpha$ can no longer be used
  to monitor the accuracy of the integration.}.

\subsection{Additional Implementation Improvements}
\label{sec:implementation}

In addition to the improvements in the coordinate system and numerical
scheme, we have made a number of additional implementation
improvements in \grayii.
One of them is the adoption of \texttt{OpenCL}, an open standard for
parallel programming\footnote{\texttt{OpenCL} was originally developed
  by Apple Inc., and currently maintained by the non-profit Khronos
  Group as an open standard.
  See \texttt{\url{https://www.khronos.org/opencl}}.}, for executing
massively parallel jobs on heterogeneous platforms.
Without any modification of the source code, \grayii\ runs on
multi-core CPUs as well as accelerators such as GPUs, Intel Xeon Phi,
and potentially Field-Programmable Gate Arrays (FPGA).

Another significant change is the adoption of the high performance
computing (HPC) framework \lux~\citep{lux} for software portability
and run time optimizations.
With \lux, the algorithms in \grayii\ are broken down into very small
modules, with multiple implementations for each of them.
This allows the users to easily construct an appropriate compilation
of modules that are specific to each application.
In addition, \lux\ benchmarks the algorithms at run time and allows
\grayii\ to automatically migrate to the most efficient algorithm.
Finally, \lux\ allows \grayii\ to use all hardware resources on a
single computing node.
It even automatically balances the work load across the CPU cores and
accelerators.

A typical application of \grayii\ is to render millions of mock images
of accretion flows onto black holes based on the output of GRMHD
simulations, for different model parameters, such as electron number
density scale and the electron-to-ion temperature ratio\citep[see,
  e.g.,][]{2015ApJ...799....1C, 2015ApJ...812..103C}.
Because a single mock image takes only a few seconds to render, there
is no need to consider inter-node communication.
Instead, millions of \grayii\ jobs can be submitted at the same time
and each job runs in parallel, independently from each other.
With this ``trivially parallelizable'' user case in mind,
\texttt{OpenCL} and \lux\ allows \grayii\ to run on a wide range of
hardware and platforms.
For example, one can compute part of the jobs on an Apple desktop,
part of the jobs on a local HPC cluster with GPUs, part of the jobs in
a supercomputing center with Xeon Phi, and the rest of the jobs in
commercial clouds such as the Amazon Web Services.
This flexibility allows us to to report benchmarks for the
implementation of (\emph{i})~different forms of the equations,
(\emph{ii})~different data structures, and
(\emph{iii})~different precisions in section~\ref{sec:benchmarks}.

\begin{figure}[b]
  \includegraphics[width=\columnwidth,trim=0 0 0 -12,clip=true]{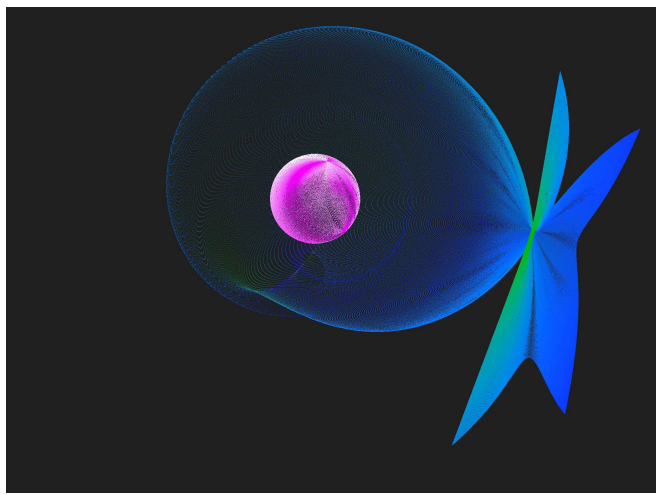}
  \caption{A screenshot of \grayii\ in its interactive mode, which
    allows a user to integrate and visualize the photon positions in
    real time.
    For this particular calculation, we set up a grid of photons
    originating at a large distance from a black hole with a spin of
    0.999.
    This grid is deformed as it passes near the black hole event
    horizon.
    Some of the photons, shown in pink here, are trapped near the
    horizon.
    Some others, shown in green and blue, are deflected at large
    angles.}
  \label{fig:fish}
\end{figure}

Figure~\ref{fig:fish} shows a screenshot of \grayii\ in its
interactive mode, which allows for the simultaneous integration and
visualization of geodesics in a black-hole spacetime.

\section{Convergence Tests}
\label{sec:tests}

In this section, we perform a number of tests with \grayii, in
situations that resemble expected realistic applications.
By default, \grayii\ uses the classic 4th-order Runge-Kutta scheme,
which is very robust and provides fast (4th-order) convergence rates.
We are interested here in its long term behavior in the Cartesian KS
coordinates \citep[see, e.g., ][for an explanation on the importance
  of long term behaviors of integrators]{2005MNRAS.364.1105S}.
Typical gravitational deflection type of tests are not useful for this
purpose because, in the Cartesian KS coordinates, the geodesic
equations are trivial at large radii.
In such cases, even though we could make the spatial interval of a
geodesic arbitrarily long, the numerical error would still be
dominated by a short segment of the geodesic near the black hole.
This does not help to monitor the long term behavior of the
integrators. Instead, we employ in our convergence tests closed,
albeit often unstable, photon and particle orbits that can be
integrated for long times.

\subsection{Unstable Spherical Photon Orbits}

Motivated by the interactive visualization by \citet{sph_orbits}, we
designed a set of tests using the unstable spherical photon orbits of
\citet{2003GReGr..35.1909T}.
These orbits are non-trivial and are excellent for observing the long
term behavior of the integrators.
While their instability may seem like a problem at first, it makes the
numerical errors accumulate (and grow) instead of canceling out and
ensures that the worst numerical scenario is explored in our tests.

\citet{2003GReGr..35.1909T} showed that any spherical orbits must lie
between the radii of the prograde and retrograde circular equatorial
orbits,
\begin{align}
  r_\mathrm{p} &\equiv 2 M \left\{1 + \cos\left[\frac{2}{3}
    \cos^{-1}\!\left(-\frac{|a|}{M}\right)\right]\right\}, \\
  r_\mathrm{r} &\equiv 2 M \left\{1 + \cos\left[\frac{2}{3}
    \cos^{-1}\!\left(\frac{|a|}{M}\right)\right]\right\},
\end{align}
which satisfy the inequalities $M \le r_\mathrm{p} \le 3 M \le
r_\mathrm{r} \le 4M$.
Therefore, our test orbits will be very close to the black hole---an
ideal place to probe the performance of the integrators.
Although no spherical orbits can pass the event horizon, some of them
can pass the ergosphere,
\begin{align}
  r_\mathrm{e} = M + \sqrt{M^2 - a^2\cos^2\theta}.
  \label{eq:ergosphere}
\end{align}
Note that we use the spherical KS coordinates $r$ and $\theta$ in the
above equations because they are equal to the BL coordinates
$\mathcal{r}$ and $\vartheta$.

\begin{deluxetable}{c|cc|cccc}
  \tablewidth{\columnwidth}
  \tablecaption{Parameters For The Convergence Study with Null
    Geodesics.
    \label{tab:tests}}
  \tablehead{Label & $a/M$ & $r/M$ & $\varPhi/M$ & $Q/M^2$ &
    \!\!\!$\max(|\cos\theta|)$\!\!\! & $\Delta\phi$}
  \startdata
\ \;A\tablenotemark{a} & 1 & 1.8 \srh      &  1.36 & 12.8304 & 0.9387 & 12.0334 \\
    B        & 1 & 2   \srh      &  1    & 16      & 0.9717 & 10.8428 \\
    C        & 1 & $1+ \sqrt{2}$ &  0    & 22.3137 & 1      &\ 3.1761 \\
    D        & 1 & $1+ \sqrt{3}$ & -1    & 25.8564 & 0.9819 & -3.7138 \\
    E        & 1 & 3   \srh      & -2    & 27      & 0.9352 & -4.0728 \\
    F        & 1 & $1+2\sqrt{2}$ & -6    &  9.6274 & 0.4634 & -4.7450 \\
    \enddata
  \tablenotetext{a}{The spherical photon orbit passes through the
    ergosphere.}
\end{deluxetable}

We list in Table~\ref{tab:tests} the parameters we use for our
convergence study.
The first column shows the label of the tests.
We consider only the extreme Kerr black hole, $a = M$, and vary the
radii of the spherical orbits.
These two input parameters are listed in the second and third columns.
With these two parameters, we can compute the normalized angular
momentum $\varPhi$ and the normalized Carter's constant $Q$ to
initialize the orbits.
For each test case, we also list, in the last two columns, the
theoretical maximum altitude, $\max(|\cos\theta|)$, that the orbit can
reach and the theoretical change in the azimuthal angle, $\Delta\phi$,
within one complete polar oscillation.
Although the spherical KS angle $\phi$ is different from the BL angle
$\varphi$, their difference depends only on $r$ (see the Appendix).
Hence, $\Delta\phi = \phi_1 - \phi_0$ and $\Delta\varphi = \varphi_1 -
\varphi_0$ are equal to each other for the same spherical orbit.


\begin{figure*}
  \begin{center}
    \includegraphics[width=2.75in,trim=36 76 72 112,clip=true]{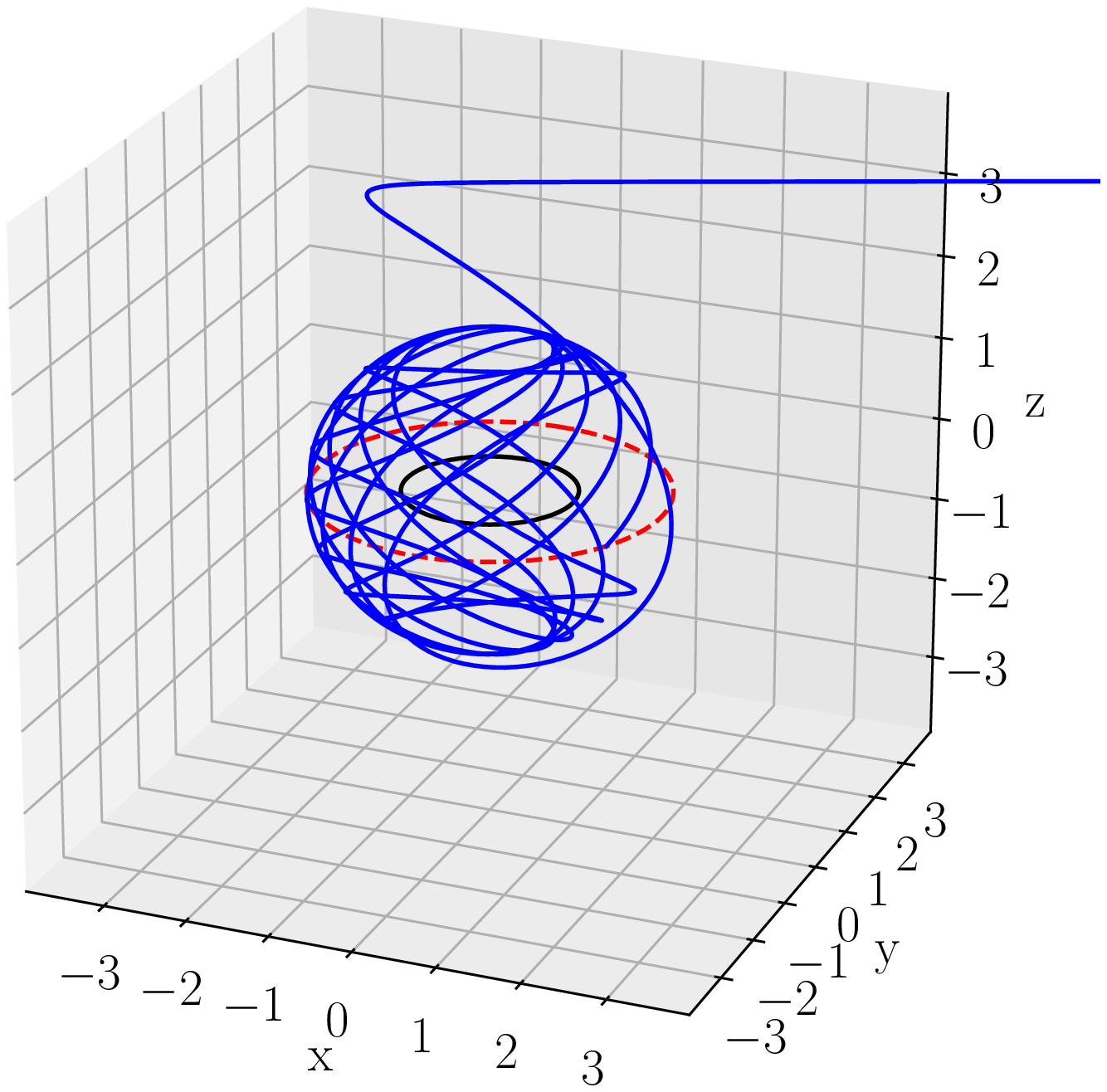}%
    \includegraphics[width=2.75in,trim=36 76 72 112,clip=true]{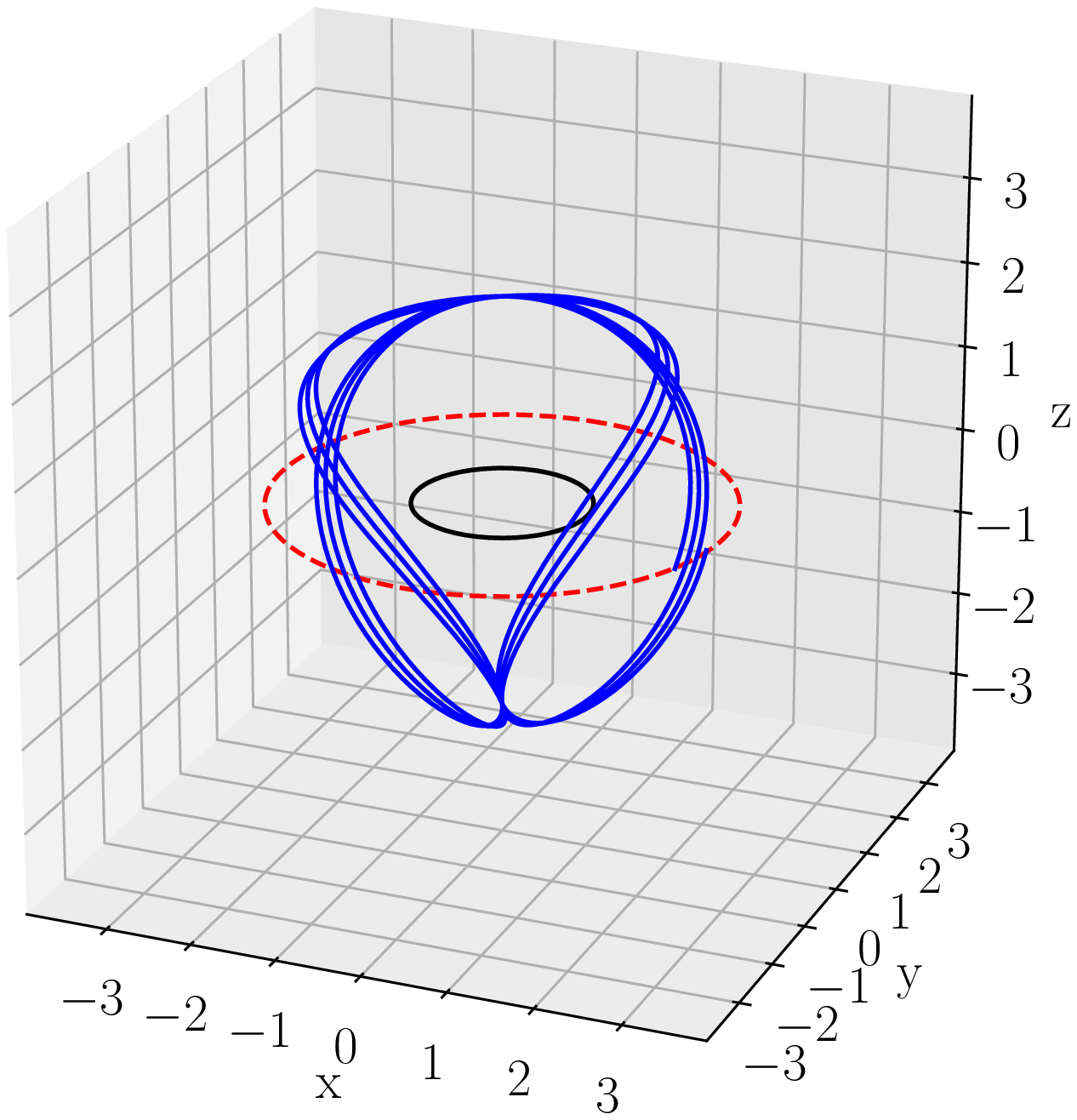}\\
    \includegraphics[width=2.75in,trim=36 76 72 112,clip=true]{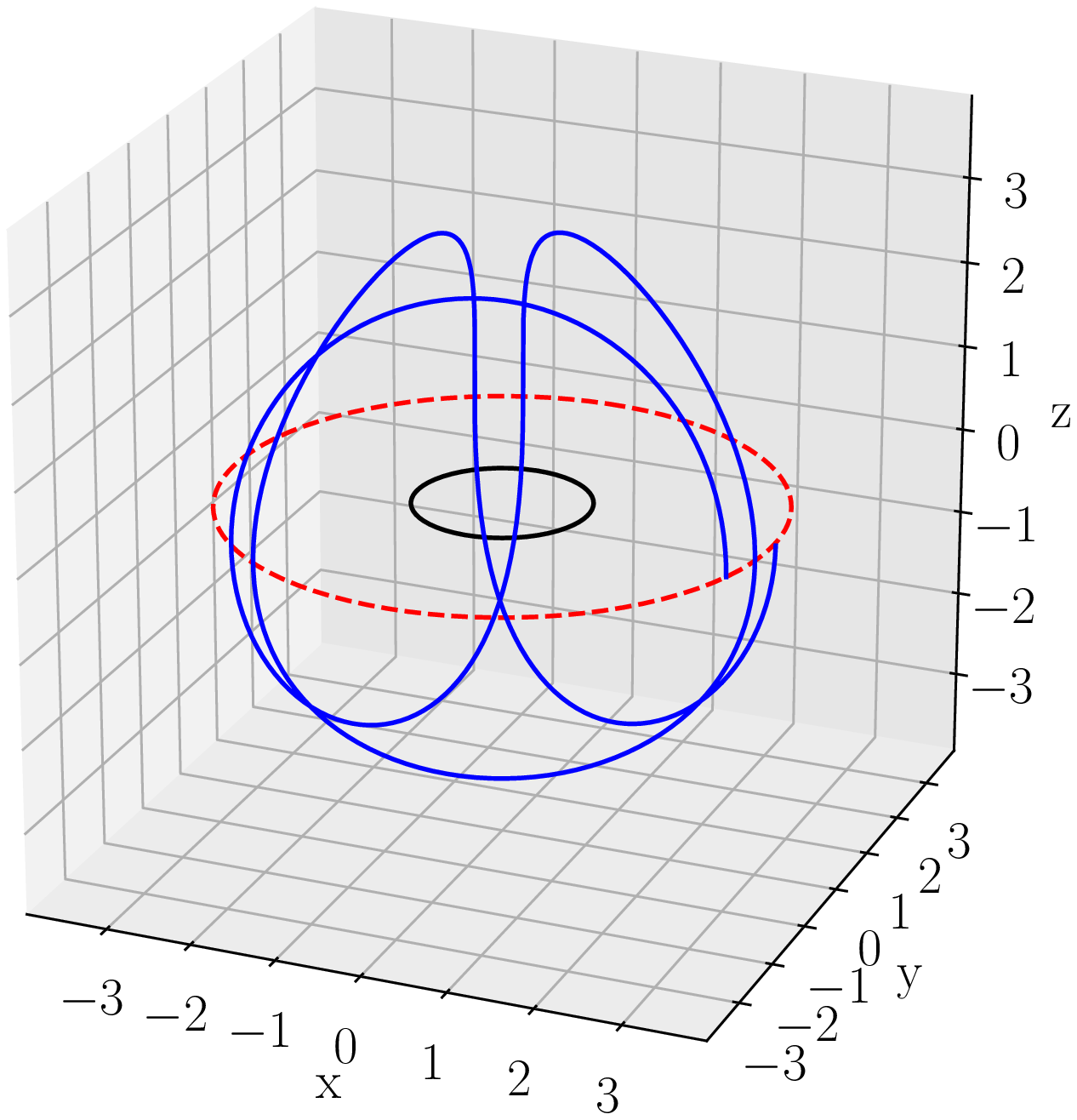}%
    \includegraphics[width=2.75in,trim=36 76 72 112,clip=true]{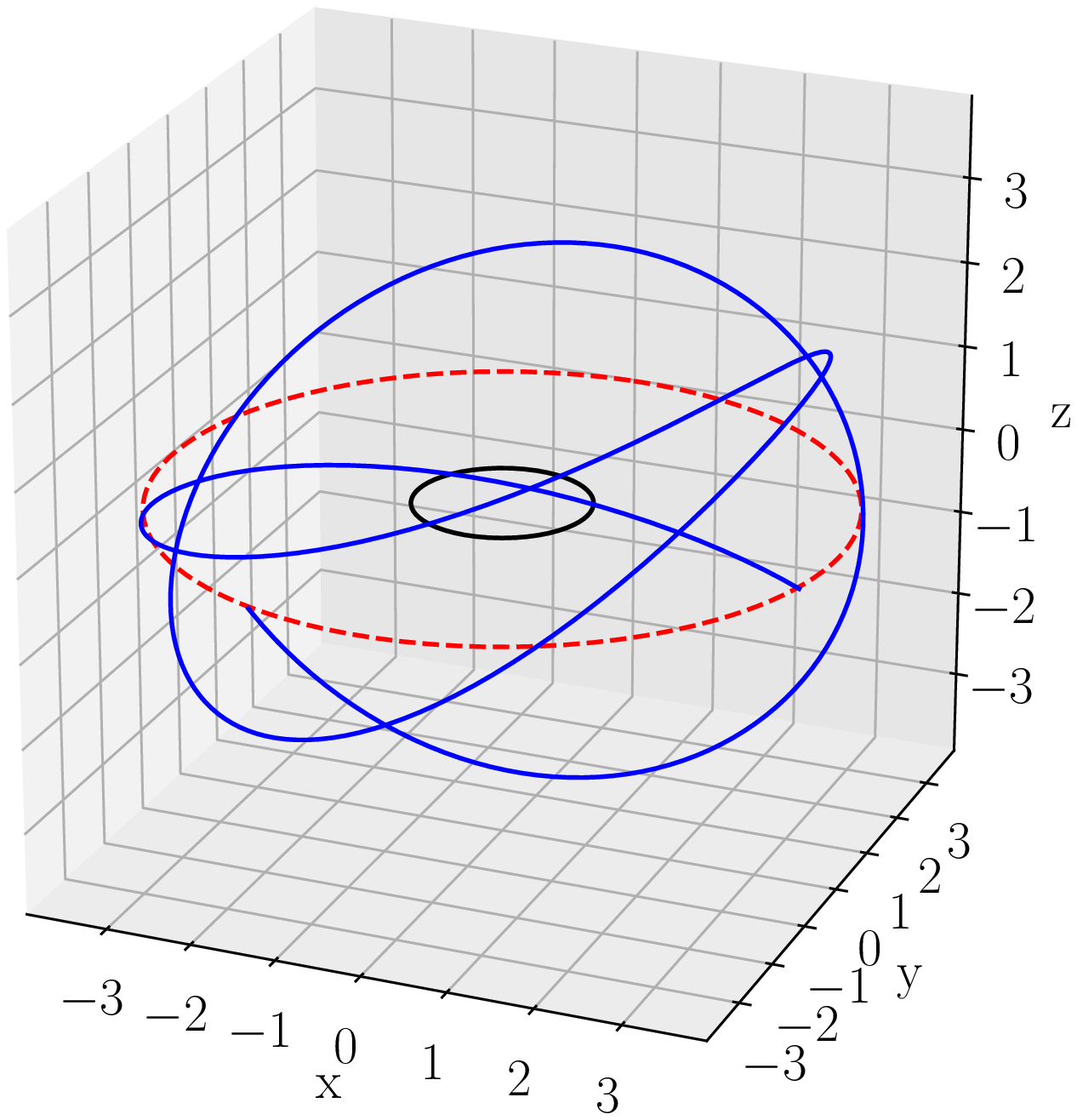}
  \end{center}
  \vspace{-16pt}
  \caption{Representative unstable spherical photon orbits around a
    black hole with $a=M$ used in our convergence study.
    (\emph{All panels}) Blue lines are the photon orbits; black solid
    circles are the (physical) singularities; and red dashed circles
    mark the radii of the spherical orbits on the equatorial planes.
    All orbits start by moving upward from the positive $x$ sides of
    the red dashed circles.
    (\emph{Top-left}) The orbit for Test~A, which passes the
    ergosphere multiple times.
    Since the orbit is \emph{unstable}, the small truncation and round
    off errors in the integrator get amplified as expected and the
    photon eventually leaves the $r = 1.8M$ sphere and flies to infinity.
    (\emph{Top-right}) The orbit for Test~C.
    This is the special case where the photon exactly passes through
    the poles.
    See Figure~\ref{fig:zoom} for a zoomed-in view of the north polar
    region.
    (\emph{Bottom-left}) The orbit for Test~E.
    The negative photon angular momentum cancels out the frame
    dragging effect exactly on the equator.
    (\emph{Bottom-right}) The orbit for Test~F.
    The initial angular momentum of the photon is negative enough
    that, unlike the other tests, the photon moves in the negative
    $\phi$ direction.}
  \label{fig:orbits}
\end{figure*}

\begin{figure}[b]
  \includegraphics[width=\columnwidth,trim=0 0 48 24,clip=false]{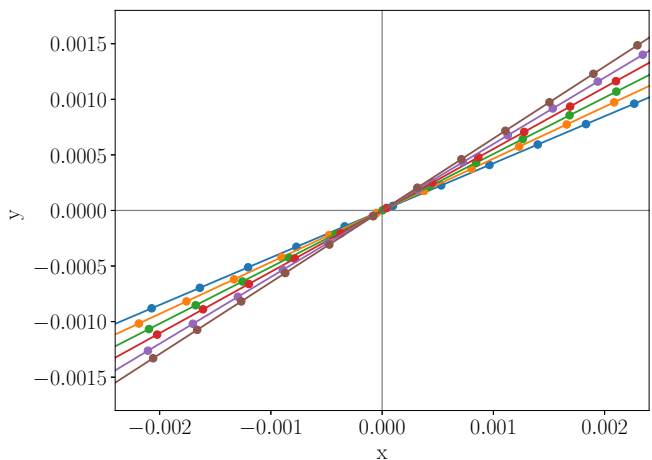}
  \vspace{-12pt}
  \caption{A zoomed-in view of the north pole of Test~A in the
    top-left panel of Figure~\ref{fig:orbits}.
    The solid circles are the actual steps of the 4th-order
    Runge-Kutta integrator with step size $\Delta\lambda = 1/1024$ and
    the color lines simply join them together for readability.
    The first pass is the blue line from top-right, the second pass is
    the orange line from bottom left, etc.}
  \vspace{12pt}
  \label{fig:zoom}
\end{figure}

Figure~\ref{fig:orbits} shows a representative set of unstable
spherical photon orbits that we used.
For all panels, the blue lines are the photon orbits; the black solid
circles are the (physical) singularities; and the red dashed circles
mark the radii of the spherical orbits on the equatorial planes.
All orbits start by moving upward from the positive $x$ sides of the
red dashed circles.

The top-left panel shows Test~A.
In this case, the radius of the orbit is small enough and the polar
momentum is large enough that the orbit oscillates in-and-out across
the ergosphere, which \grayii\ in Cartesian KS has no problem
handling.
Note that, since the orbit is \emph{unstable} and we do not put any
constraints on the integrator in \grayii, the small truncation and
round off errors in the integrator get amplified as expected.
Because of the accumulation of these numerical errors, the photon
eventually leaves the $r = 1.8M$ sphere and flies to infinity.

The top-right panel shows Test~C, for which the whole orbit is now
outside the ergosphere.
Although the photon does not have any angular momentum in this case
($\varPhi = 0$), the orbit is tilted on the equatorial plane because
of frame dragging.
Also, this is the special case for which the photon exactly passes the
poles multiple times at $x = y = 0$ and $z \approx \pm 2.41$.
In Figure~\ref{fig:zoom} we zoom into the north polar region of
Test~C.
The solid circles are the actual steps of the 4th-order Runge-Kutta
integrator with step size $\Delta\lambda = 1/1024$ and the colored
lines simply join them together for clarity.
The first pass is the blue line from lower-left; the second pass is
the orange line from top-right; etc.
Again, the integrator has no problem handling the pole because there
is no coordinate singularity in the Cartesian KS coordinates.

In the bottom-left panel, we plot the unstable spherical photon orbit
for Test~E.
Although the photon angular momentum is negative, it cancels out the
frame dragging effect exactly on the equatorial plane so the photon
moves vertically when it passes the equator.
In the bottom-right panel, we plot the photon orbit for Test~F.
This time, the initial angular momentum is negative enough that the
photon finally moves in the negative $\phi$ direction.

\begin{figure*}
  \begin{center}
    \includegraphics[width=\columnwidth,trim=0 0 48 24,clip=true]{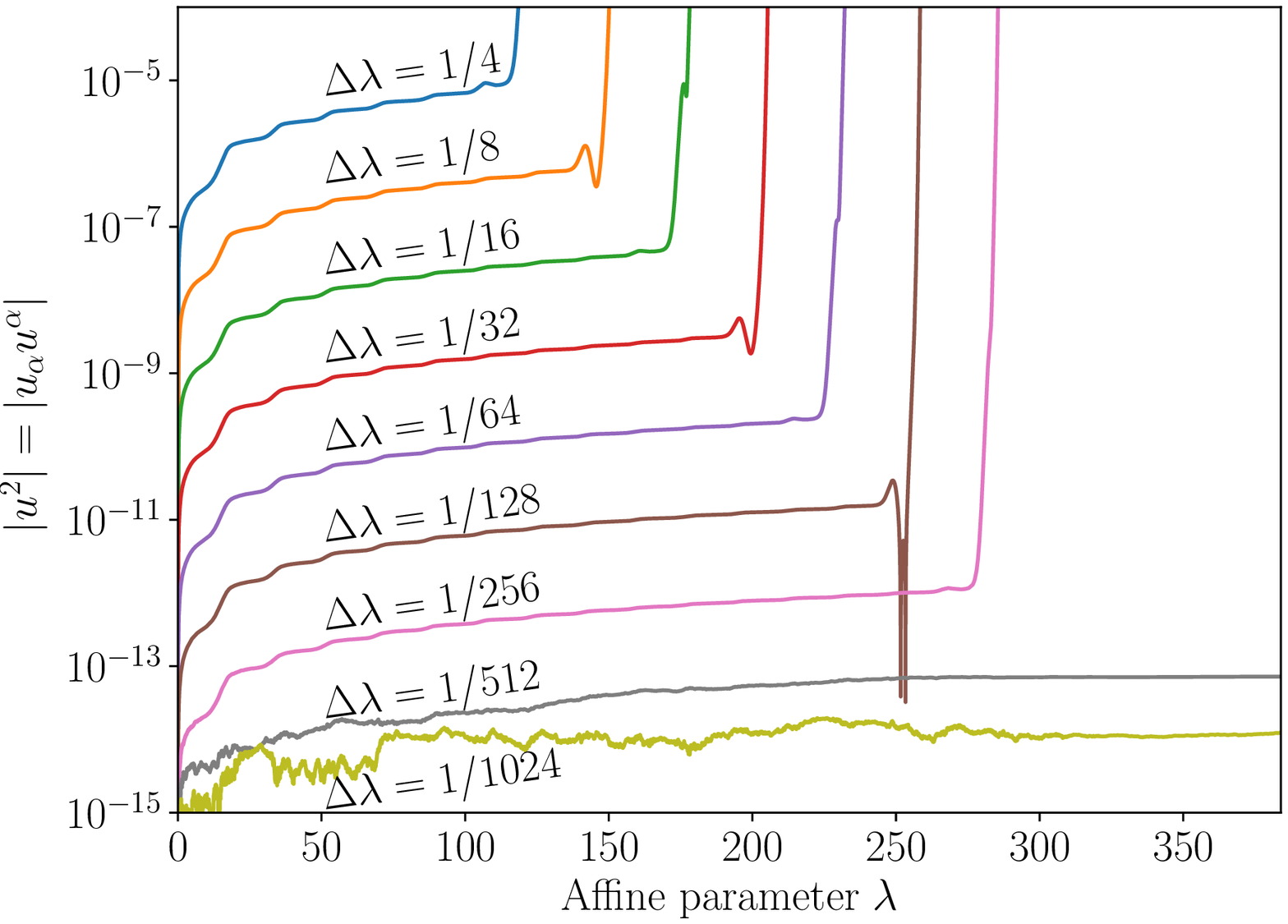}%
    \includegraphics[width=\columnwidth,trim=0 0 48 24,clip=true]{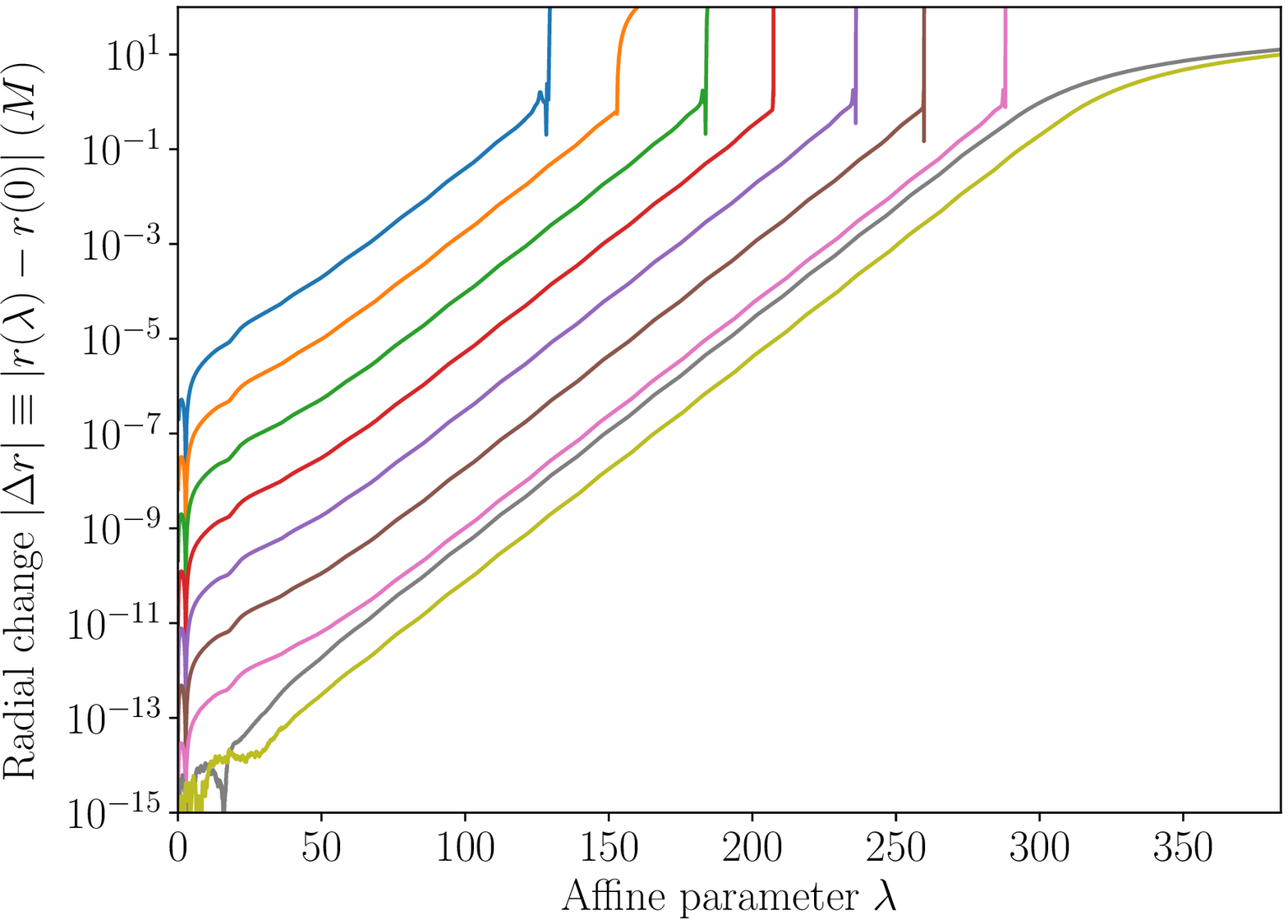}
  \end{center}
  \vspace{-12pt}
  \caption{(\emph{Left}) The time evolution of $u^2 = u_\alpha
    u^\alpha$ for Test~A.
    We use nine different step sizes in this test, namely,
    $\Delta\lambda = 1/4$, 1/8, ..., 1/512, and 1/1024, which are
    labeled in the left panel.
    For all step sizes, \smash{$u^2$} is initially of order
    \smash{$10^{-16}$} and increases approximately linearly until the
    numerical solutions blow up.
    (\emph{Right}) The change of the radius of the photon orbit as a
    function of $\lambda$.
    The color coding of the curves follows that of the left panel.}
  \label{fig:evolr}
\end{figure*}

In Figure~\ref{fig:evolr}, we plot two of the error indicators for
Test~A.
The left panel shows the time evolution of $u^2 = u_\alpha u^\alpha$.
It is a constant of motion and should remain zero for photons.
In \grayii, because we integrate the geodesic equations without
explicitly using any constants of motion, $u^2$ is, in principle,
unconstrained in the numerical integration.
Therefore, its value is a good measure of the numerical errors.
We use nine different step sizes in this test, covering the range
$\Delta\lambda = 1/4$, 1/8, ..., 1/512, and 1/1024, which are labeled
on the left panel.
For all step sizes, $u^2$ is initially of order $10^{-16}$ and
increases approximately linearly until the numerical solutions blow
up.
Clearly, a smaller step size leads to smaller \smash{$u^2$} and,
hence, smaller error.

To understand why the numerical solutions blow up, we plot the changes
of the radius of the photon orbit as a function of $\lambda$ in the
right panel.
Recalling that the tests are for unstable spherical photon orbits,
$|\Delta r| \equiv |r(\lambda) - r(0)|$ should grow exponentially,
with an amplitude proportional to the perturbation---or to the
numerical errors in our case---of the orbits.
The coloring of the curves is identical to the left panel, namely,
blue is for $\Delta\lambda = 1/4$, orange is for $\Delta\lambda =
1/8$, etc.
All the curves grow as expected and blow up at around $|\Delta r| \sim
1$.
This is not a coincidence.
In fact, for all orbits except the gray and yellow ones, the photons
approach the physical singularities as they depart from their
spherical orbits.
When they get very close to the singularities, the fixed time steps
fail to integrate the geodesics, resulting in significant jumps in
both $u^2$ and $\Delta r$.
For the gray and yellow curves, the photons actually move away from
the singularities as they depart from the spherical orbits.
Therefore, although the photons approach infinity (linearly),
\smash{$u^2$} always remain small.

\begin{figure}[b]
  \includegraphics[width=\columnwidth,trim=0 0 48 24,clip=true]{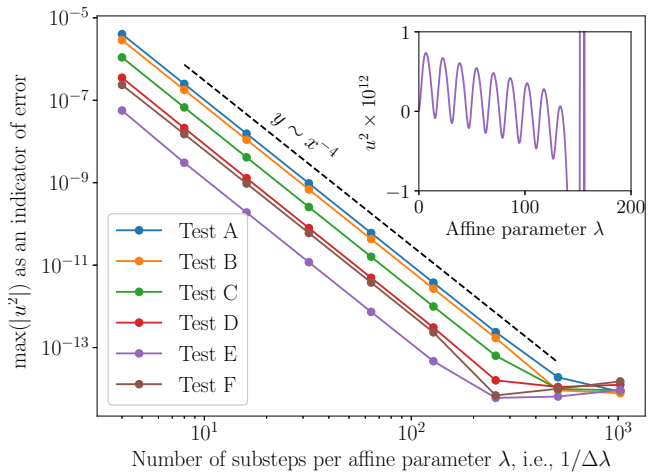}
  \vspace{-12pt}
  \caption{Results of the convergence study using the integral of
    motion $u^2$. Here, we plot $\max(u^2)$ for $0 \le \lambda < 64$
    for the six test problems we performed, as a function of the step
    $\Delta\lambda$ in the affine parameter.
    \grayii\ converges at 4th order it used the 4th-order Runge-Kutta
    scheme.
    The amplitude of the error decreases as $r$ increases except for
    Test~E.
    This is an artifact of the oscillatory behavior of $u^2$, in this
    test, shown in the inset.}
  \label{fig:uu-conv}
\end{figure}

In Figure~\ref{fig:uu-conv}, we summarize the convergence properties
of our algorithm as quantified by the integral of motion $u^2$.
We plot $\max(|u^2|)$ for $0\le\lambda\le 64$ for all six test
problems.
For each test problem, we change $\Delta\lambda$ in the range 1/4,
1/8, 1/16, ..., to 1/1024.
As Figure~\ref{fig:evolr} already showed, the errors decrease as we
use smaller step sizes and \grayii\ converges at 4th order---an
expected result because of the 4th-order Runge-Kutta scheme we are
using.

\begin{figure*}
  \includegraphics[width=\columnwidth,trim=0 0 48 24,clip=true]{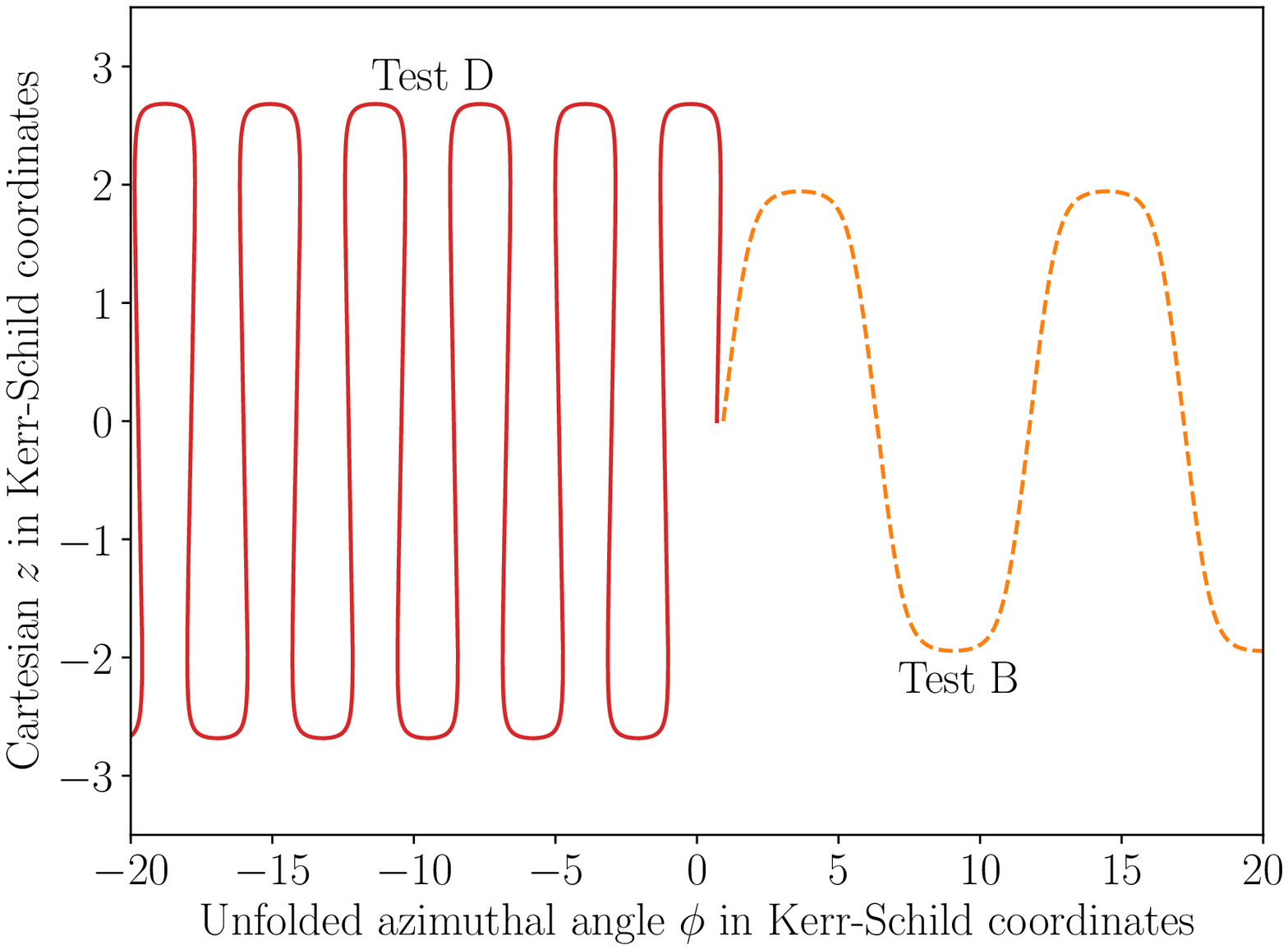}%
  \includegraphics[width=\columnwidth,trim=0 0 48 24,clip=true]{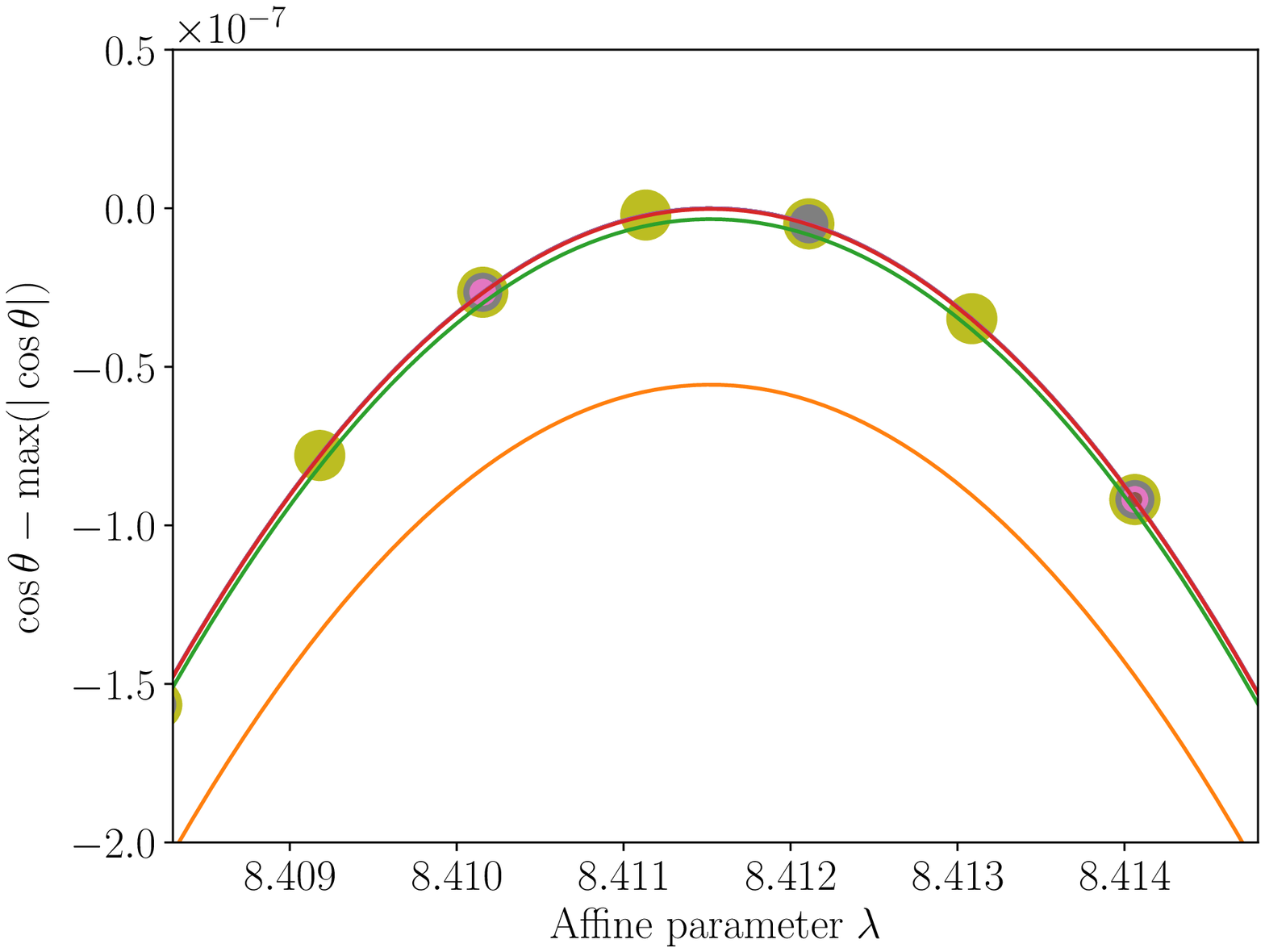}
  \caption{(\emph{Left}) Unfolded photon orbits in a cylindrical
    coordinate projection.
    The vertical axis is the Cartesian $z$ in the Kerr-Schild
    coordinates.
    The horizontal axis is the unfolded azimuthal angle $\phi$ in the
    Kerr-Schild coordinates.
    We only plot two representative curves here for Tests~B and D.
    (\emph{Right}) Zoom-in view of the first peak of Model~B as a
    function of the affine parameter $\lambda$ with different step sizes.
    The solid circles are the actual output of the \grayii.
    For each step size, we fit a quadratic equation to the largest
    three circles.
    The quadratic equations are then plotted as the different curves
    here.
    The color scheme of the solid circles and the curves matches those
    of Figure~\ref{fig:evolr}.}
  \label{fig:z-phi}
\end{figure*}

For a more detailed test of the geodesics, in the left panel of
Figure~\ref{fig:z-phi}, we unfold the photon orbits in the azimuthal
direction and plot the coordinate $z$ as a function of $\phi$.
To avoid overlap between the oscillatory curves, we plot the orbits
only for Test~B and D.
The photon in Test~B has positive angular momentum, hence it moves
toward positive $\phi$.
In the same figure, the photon in Test~D has a small negative angular
momentum.
Although its overall orbit points toward negative $\phi$, the photon
changes its direction near the equator because of the stronger frame
dragging effect there.

\begin{figure}[b]
  \includegraphics[width=\columnwidth,trim=0 0 48 24,clip=true]{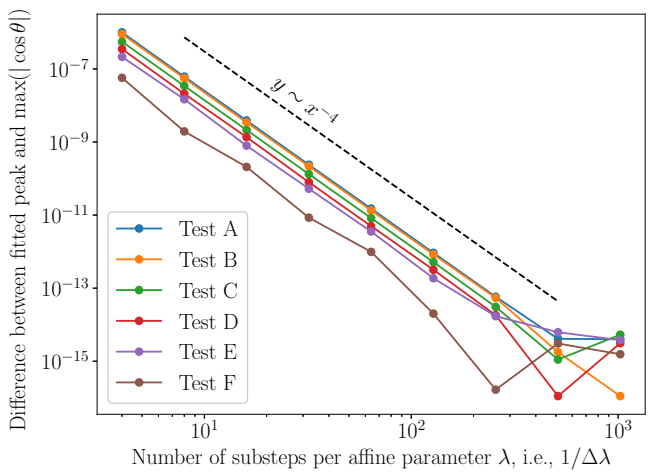}
  \vspace{-12pt}
  \caption{Results of the convergence study using the maximum polar
    angles a photon reaches during its orbit.
    The plot is similar to Figure~\ref{fig:uu-conv}.
    It shows that the difference between the fitted peak value of the
    numerical solution converges to the analytical one at the expected
    4-th order.}
  \label{fig:z-conv}
\end{figure}

The peak values of the $z-$coordinate in the left panel of
Figure~\ref{fig:z-phi} can be calculated analytically, are given in
equation~(8) in \citet{2003GReGr..35.1909T} and listed in the sixth
column of our Table~\ref{tab:tests}.
We can use these as a second convergence test of \grayii.
(Note that, for convenience, we plot in the following figures the
maximum cosine of the polar angle, i.e., $\max\vert\cos\theta\vert$,
instead of the maximum $z-$coordinate of each photon orbit).
The numerical values at the local maxima depend strongly on the
resolution because of sampling effect.
This is illustrated in the right panel of Figure~\ref{fig:z-phi},
where we zoom into the first peak of Model~B as a function of the
affine parameter $\lambda$ with different step sizes.
The solid circles are the actual steps of the 4th-order Runge-Kutta
scheme.
It is clear that the circles are offset from the locations of the
peaks.
To overcome this sampling effect in polluting our convergence test, we
fit a quadratic equation to the largest three points for each
resolution.
The curves in the right panel of Figure~\ref{fig:z-phi} correspond to
these quadratic equations.
From the plot, even the red curve with $\Delta\lambda = 1/32$ is
visually indistinguishable from the more accurate curves at this
scale.
Only at $\Delta\lambda = 1/16$, the green curve starts to deviate from
the more accurate curves.
We compute the differences between the peak values of the fitted
quadratic equations and the analytical values.
The results are plotted in Figure~\ref{fig:z-conv}, which shows again
a 4th-order convergence rate for all the tests.

\begin{figure}[b]
  \includegraphics[width=\columnwidth,trim=0 0 48 24,clip=true]{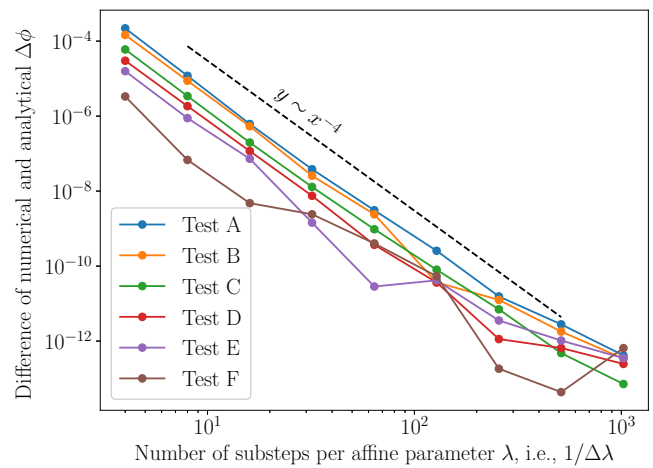}
  \vspace{-12pt}
  \caption{Results of the convergence study using the change in the
    azimuth of the photon orbit during one complete oscillation in
    latitude.
    The plot is similar to Figure~\ref{fig:uu-conv} and
    \ref{fig:z-conv}.
    It shows that the difference between the numerical and analytic
    values of $\Delta\phi$ converges at the expected 4-th order.}
  \label{fig:dphi-conv}
\end{figure}

Our final convergence test uses another result from
\citet{2003GReGr..35.1909T}, who derived the equation to integrate the
change in azimuth for one complete oscillation in latitude.
Although there is still a sampling effect due to the change of step
size, its resolution is much simpler.
We simply use linear interpolation to obtain the root of $z(\phi)$ at
the first complete cycle and subtract the initial coordinate $\phi$
from it.
This numerical value is then compared with the numerical integration
of equation~(17) in \citet{2003GReGr..35.1909T}.
The final result is plotted in Figure~\ref{fig:dphi-conv}, which again
shows a 4th-order convergence rate for all the tests.

\subsection{Stable Circular Particle Orbits}

\begin{figure}[b]
  \includegraphics[width=\columnwidth,trim=0 0 48 24,clip=true]{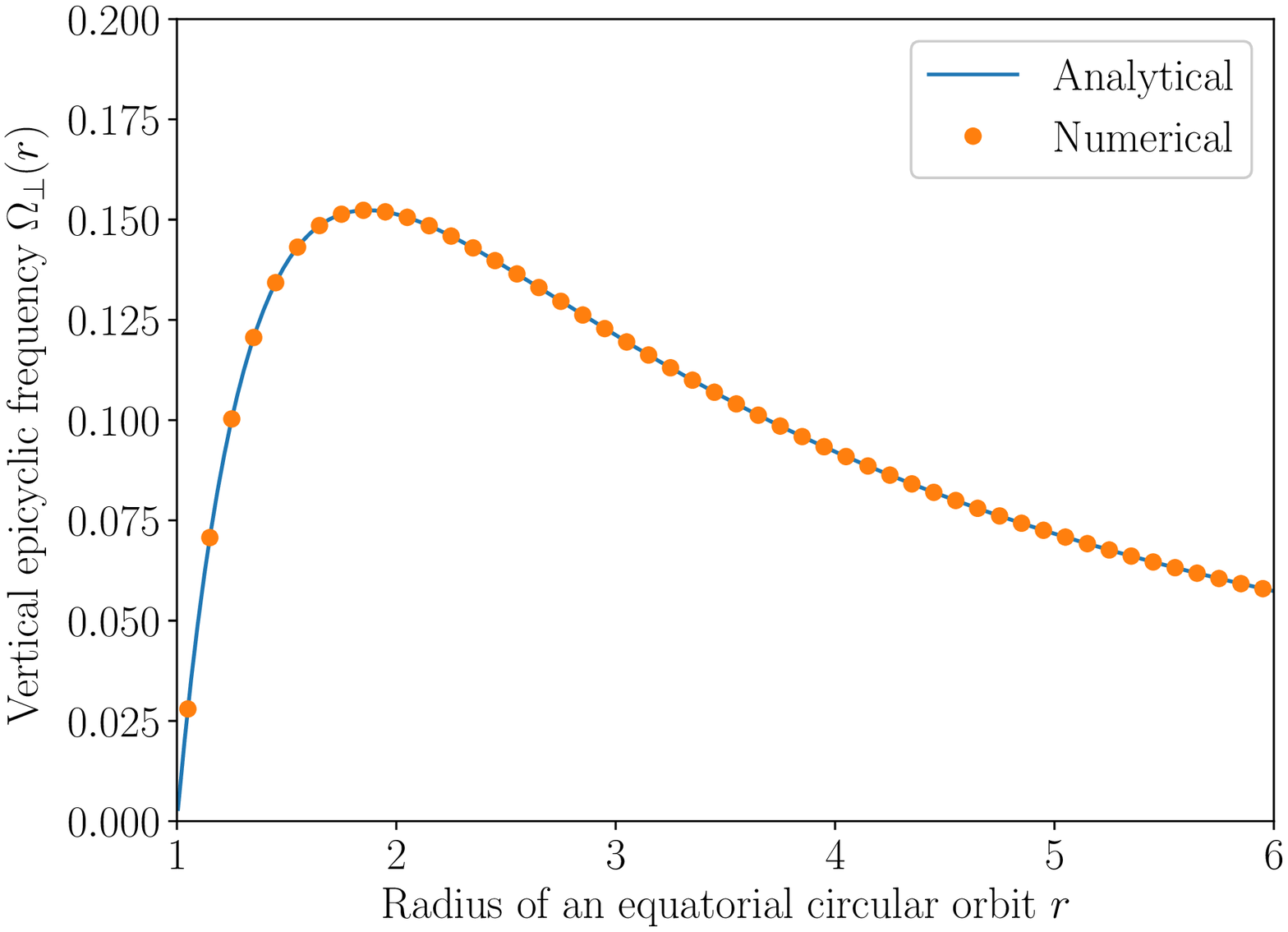}
  \vspace{-12pt}
  \caption{Vertical epicyclic frequency as a function of orbital
    radius.
    We use \grayii\ to integrate a nearly circular time-like geodesics
    around a black hole of spin $a = 1$.
    We introduce a small initial vertical velocity $v_z = 10^{-12}$ to
    induce vertical oscillation and precession.
    We then compare the numerical vertical epicyclic frequencies
    (orange circles) with its analytical expression (blue solid line).
    The fractional difference between the two curves at a radius
    $r<1.5$ is less than $10^{-13}$.}
  \label{fig:OmegaV}
\end{figure}

Since \grayii\ makes no specific assumption about geodesics, it can
integrate orbits for massive test particles, i.e., time-like
geodesics, without any modification of the integrator.
As a second set of convergence tests, we integrate stable, nearly
circular orbits at different radii around a black hole with a spin of
$a = 1$.

Nearly circular orbits in general relativity precess because of the
deviation of the effective gravitational potential from the Newtonian
$1/r$ form.
One can describe completely their motion along the three polar
coordinates using three independent oscillatory frequencies, one
azimuthal (the orbital frequency $\Omega$), one radial (the radial
epicyclic frequency $\kappa$), and one vertical (the vertical
epicyclic frequency $\Omega_\perp$).
The radial epicyclic frequency vanishes at the location of the
innermost stable circular orbit and becomes imaginary inside that
radius.
In the same region, the vertical epicyclic frequency is significantly
smaller than the orbital frequency.
All three frequencies for a nearly circular orbit can be calculated
analytically \citep[see, e.g.,][]{2008ApJ...680.1319S}.

In order to force a small precession of the orbital plane of a test
particle, we introduce a small vertical velocity, $v_z = 10^{-12}$, to
the initial conditions that would otherwise lead to a circular orbit.
We then use the numerical orbit to calculate the vertical epicyclic
frequency (as measured by an observer at infinity) and compare it to
the analytic expression.
In order to avoid the numerical effects described in the previous
section, we perform linear interpolations between the calculated
points in $z(t)$ and measure the time interval between successive
maxima.

In Figure~\ref{fig:OmegaV}, we overplot the numerical measurement from
\grayii\ on the analytic expression, as a function of the radius of
the orbit in BL coordinates.
The numerical and analytical results are indistinguishable.
In fact, for $r \lesssim 1.5$, the difference between the numerical
and analytical results is less than $10^{-13}$.
For all other convergence properties related to the test-particle
orbits, we found no appreciable difference with the results shown
earlier for the null geodesics.

\section{Profiling and Benchmarks}
\label{sec:benchmarks}

\begin{deluxetable*}{ccc|cccccccc}
  \tablecaption{Benchmark results for integration of null
    geodesics\tablenotemark{a}
    \label{tab:elapse}}
  \tablehead{Precision & Coordinates & Data Order &
    i7-3720QM\tablenotemark{b} & E5-2650$\times2$\tablenotemark{c} &
    HD4000 & GT650M & M2090 & K20X & GTX780 & Titan Black}
  \startdata
  Single & BL & \ \,AoS\tablenotemark{d}           & 43.1 & 6.07 & 2.99 & 2.49 & 0.597 & 0.223 & 0.174 & 0.167 \\
  Single & BL & \smash{\ \ \,SOA\tablenotemark{e}} & 49.4 & 6.01 & 2.98 & 2.89 & 0.597 & 0.219 & 0.170 & 0.175 \\
  Single & KS & AoS                                & 57.5 & 6.27 & 4.53 & 2.69 & 0.995 & 0.420 & 0.311 & 0.297 \\
  Single & KS & SoA                                & 54.8 & 6.20 & 4.55 & 2.31 & 0.995 & 0.422 & 0.315 & 0.309 \\
  Double & BL & AoS                                & 55.4 & 16.6 & ---  & 70.6 & ---   & 3.90  & 6.44  & 1.65  \\
  Double & BL & SoA                                & 56.7 & 16.2 & ---  & 80.6 & ---   & 3.90  & 6.44  & 1.66  \\
  Double & KS & AoS                                & 66.0 & 17.4 & ---  & 59.1 & ---   & 2.41  & 6.95  & 1.15  \\
  Double & KS & SoA                                & 63.1 & 17.5 & ---  & 59.0 & ---   & 2.40  & 6.96  & 1.15  \\
  \enddata
  \tablenotetext{a}{The elapsed time per single 4th-order Runge-Kutta
    time step of a single ray in \emph{nanoseconds}.
    Smaller values are better.}
  \tablenotetext{b}{i7-3720QM is a 4-core mobile CPU}
  \tablenotetext{c}{E5-2650 is an 8-core server CPU and there are
    two CPUs per node on the \texttt{El~Gato} supercomputer at the
    University of Arizona}
  \tablenotetext{d}{Array-of-Structures}
  \tablenotetext{e}{Structure-of-Arrays}
\end{deluxetable*}

In section~\ref{sec:KS}, we reduced the operation-count of the
geodesic equations in the Cartesian KS coordinates by a series of
mathematical manipulations.
This suggests that, theoretically, solving this optimized form of
geodesic equations is not much more expensive than in the BL
coordinates.
In this section, we look at the actual benchmarks on different devices
to support our assertions.

The most direct method to compare the performances of the geodesic
equations in the Cartesian KS and BL forms is to look at the elapsed
time for a single 4th-order Runge-Kutta time step of a single ray,
which contains four evaluations of the RHS of the geodesic equations.
However, modern accelerators such as GPUs are effectively vector
processors.
They are only efficient when a large number of threads are executed in
parallel.
In addition, these accelerators are not general purpose devices.
Driving them requires a host---usually a full power CPU---to compile
the kernels and send instructions and data to the devices.
These overheads can sometimes be quite significant compared to the
computations.

\begin{deluxetable*}{ccc|cccccccc}
  \tablecaption{Benhcmarks Results for Null Geodesics\tablenotemark{a}
    \label{tab:speedup}}
  \tablehead{Precision & Coordinates & Data Order & i7-3720QM &
    E5-2650$\times$2 & HD4000 & GT650M & M2090 & K20X & GTX780 & Titan Black}
  \startdata
  Single & BL & AoS & 2.25 & 16.0 & 32.5 & 39.0 & 163  & 436  & 558  & 582 \\
  Single & BL & SoA & 1.95 & 16.0 & 32.3 & 33.3 & 161  & 439  & 566  & 549 \\
  Single & KS & AoS & 1.74 & 16.0 & 22.2 & 37.3 & 101  & 239  & 323  & 338 \\
  Single & KS & SoA & 1.81 & 16.0 & 21.8 & 42.9 & 99.7 & 235  & 315  & 321 \\
  Double & BL & AoS & 4.79 & 16.0 & ---  & 3.76 & ---  & 68.1 & 41.2 & 161 \\
  Double & BL & SoA & 4.57 & 16.0 & ---  & 3.22 & ---  & 66.5 & 40.3 & 156 \\
  Double & KS & AoS & 4.22 & 16.0 & ---  & 4.71 & ---  & 116  & 40.1 & 242 \\
  Double & KS & SoA & 4.44 & 16.0 & ---  & 4.75 & ---  & 117 & 40.2 & 243  \\
  \enddata
  \tablenotetext{a}{Same benchmark results as in
    Table~\ref{tab:elapse} but in terms of speedup using a single core
    of E5-2650 as the baseline.
    Because there are two E5-2650 CPUs per node on \texttt{El~Gato},
    we \emph{define} its speedup to 16.
    All other values are scaled accordingly.
    Unlike Table~\ref{tab:elapse}, larger values are better.}
\end{deluxetable*}

We design our benchmarks to reduce the impacts of the above factors.
Following the approach found in a typical application, we use
\grayii\ to integrate backwards null geodesics from an image at an
initial radius $r = 1024M$ toward a black hole of spin $a = 0.999M$.
The integration is performed by executing the same \texttt{OpenCL}
kernel $\sim 64$ times, until all photons pass by or are close enough
to the event horizon, or escape to distances larger than the initial
radius.
The kernel performs host-to-device communication once right after it
starts the 4th-order Runge-Kutta methods with fixed step size 1024
times.
It also performs device-to-host communications once more just before
it ends.
We measure the elapsed time between the instants when the kernel
starts and ends.
Typically, the elapsed time is reduced a bit after the first few
kernel executions due to instruction loading and caching and then
saturates toward a steady value.

Because the performance of GPUs is sensitive to how the computations
are grouped and distributed to different sub-processors (for CPUs they
are called ``cores''; for nVidia GPUs they care called
``multiprocessors''), in order to measure the peak performance, we
repeat the above process with five different resolutions of images:
$64\times64$, $128\times128$, $256\times256$, $512\times512$, and
$1024\times1024$, and allow \texttt{lux}'s run time performance tuning
algorithm to choose the optimal workgroup size.
We compute the elapsed time of a single step for each ray by dividing
the shortest measured time by 1024 and the total number of rays.

We list the benchmark results in Table~\ref{tab:elapse}.
The first three columns are the precision, the coordinate system, and
the data-order used in \grayii, respectively.
The numbers in all other columns are in nanoseconds.
The 4th column is for a mobile/laptop 4-core CPU and the 5th column is
for two 8-core server CPUs (i.e., 16 cores in total).
The 6th column is for an Intel integrated graphics chip.
The 7th--11th columns are for different nVidia GPUs.

All the tested processors, except Intel HD4000, support both single
and double precisions.
For nVidia's Tesla M2090, our workstation failed to compile \grayii's
\texttt{OpenCL} kernel---this may be due to the limitation of the
driver or a bug in the \texttt{OpenCL} implementation.
For CPUs,
the performance of single precision is only slightly (or a factor of a
few) faster than for double precision.
For the mobile and consumer graphics chips GT650M and GTX780, single
precision is significantly faster---up to a factor of $\sim 28$---than
double precision.
This is no surprise.
Single precision operations are good enough for computer graphics and
gaming applications.
Hence, a large number of transistors on these consumer graphics chips
are used to perform single precision operations only.
For the HPC specific GPU Tesla K20X and high-end graphics cards nVidia
Titan Black, the performance difference between single and double
precisions is less dramatic.

For single precision, integrating in the KS coordinates is slightly
more expensive than in the BL coordinates but the difference is
usually less than $\sim 70\%$ (except for K20X).
This supports our estimate in section~\ref{sec:KS} that integrating in
KS coordinates has roughly 65\% more operations compared to
integrating in BL coordinates.
For double precision, it is interesting to note that, for most GPUs
(except GTX780), KS integration is actually faster than BL integration
by up to $\sim 40\%$.
There are two main reasons for this.
First, BL coordinates require evaluations of trigonometric functions,
which are expensive in double precision.
Second, the equations in KS are highly symmetric, which allows the
compiler to optimize them by hardware-accelerated vector instructions.

Finally, there is no significant difference between the two different
approaches to ordering the structures and arrays for these benchmarks.
This is just an indication that our benchmarks successfully overcome
memory access overhead and are able to reveal the actual computation
performance.

In order to more easily read off the performance gain, we convert the
elapsed time in Table~\ref{tab:elapse} to speedups in
Table~\ref{tab:speedup}.
Larger numbers mean higher speedups.
We use a single core of the E5-2650 CPU as our baseline.
Hence, two E5-2650 CPUs give us 16 cores in the 5th column.
For single precision, most of the GPUs are $100-600$ times faster than
a single E5-2650 core.
For double precision, although the speedups are not as large, the HPC
specific Tesla K20X and high-end graphics cards nVidia Titan Black are
still two orders of magnitude faster than a single E5-2650 core.

\section{Summary}
\label{sec:summary}

In this paper, we presented \grayii, a new open source,
hardware-accelerated, general relativistic integrator for time-like
and null geodesics.
By using the Cartesian form of Kerr-Schild coordinates, integration in
\grayii\ avoids all coordinate singularities at the pole and at the
horizon that are present in Boyer-Lindquist coordinates.
Using a rearranged form of the geodesic equations (see
equations~[\ref{eq:opt0}] and [\ref{eq:opti}]) makes the integrator in
Cartesian KS coordinates as efficient as in BL coordinates.
In addition, by using the \texttt{OpenCL} standard and the HPC
framework \texttt{lux}, \grayii\ runs optimally on a wide range of
software platforms and hardware devices.

We carefully examined the properties of the numerical algorithm and
showed that, for a number of different problems with known analytic
solutions, it converges at the expected rate.
We also report extensive performance benchmarks and show the
significant (1-2 orders of magnitude) improvement in the efficiency of
\grayii\ when it is run on GPU architectures.

This algorithm is optimally suited for massively parallel integration
of geodesics.
It can be utilized for computing the transport of radiation through
black hole accretion flows, the evolution of a cluster of particles in
the vicinity of a black hole, or even direct particle or gyrokinetic
simulations of plasmas in Kerr spacetimes.

\acknowledgements

C.K.C., D.P., and F.O. are partially supported by NASA TCAN award
NNX~14AB48G and NSF grant AST~1312034; L.M. is supported by NFS GRFP
grant DGE~1144085.
F.O. gratefully acknowledges a fellowship from the John Simon
Guggenheim Memorial Foundation.
D.P. acknowledges support from the Radcliffe Institute for Advanced
Study at Harvard University.
All authors acknowledge the hospitality of the Black Hole Initiative
at Harvard University, which is supported by a grant from the John
Templeton Foundation.
The ray tracing calculations were performed on the \texttt{El~Gato}
GPU cluster, the \texttt{tesla} and \texttt{merope} GPU workstations
at the University of Arizona, and C.K.C.'s laptop.
\texttt{El~Gato} is funded by NSF award 1228509; \texttt{tesla} is
funded by the theoretical astrophysics program at Arizona; and
\texttt{merope} is provided by Andras Gaspar with nVidia's hardware
donation.
We also thank Ramesh Narayan and Jonathan McKinney for valuable
comments and discussions.


\bibliography{ms,my}

\end{document}